\documentclass[prl,twocolumn,groupedaddress]{revtex4-1}
\usepackage{graphicx}
\usepackage{dcolumn}
\usepackage{bm}
\usepackage{amsmath}
\usepackage{graphicx}
\usepackage{color}
\usepackage{wrapfig}
\usepackage{epstopdf}

\begin{document}

\title{Ultrafast Magnetic Light}

\author{Sergey V. Makarov$^{1}$, Sergey I. Kudryashov$^{1,2}$ , Alexander E. Krasnok$^{1}$, and Pavel A. Belov$^{1}$}

\address{$^{1}$ITMO University, St.~Petersburg 197101, Russia\\
$^{2}$Lebedev Physical Institute, Moscow 119991, Russia}

\begin{abstract}
  We propose a novel concept for efficient dynamic tuning of optical properties of a high refractive index subwavelength nanoparticle with a magnetic Mie-type resonance by means of femtosecond laser radiation. This concept is based on ultrafast generation of electron-hole plasma within such nanoparticle, drastically changing its transient dielectric permittivity. This allows to manipulate by both electric and magnetic nanoparticle responses, resulting in dramatic changes of its extinction cross section and scattering diagram. Specifically, we demonstrate the effect of ultrafast switching-on a Huygens source in the vicinity of the magnetic dipole resonance. This approach enables to design ultrafast and compact optical switchers and modulators based on the "ultrafast magnetic light" concept.
\end{abstract}

\maketitle

\textbf{Introduction}--All-dielectric "magnetic light" nanophotonics based on nanoparticles of high refractive index materials allows manipulation of a magnetic component of light at nanoscale without high dissipative losses, inherent for metallic nanostructures~\cite{Cummer_08, Zhao09, evlyukhin2010, kuznetsov2012, Evlyukhin:NL:2012, Miroshnichenko:NL:2012, Brener_12,  krasnok2015towards}. This "magnetic light" concept has been implemented for nanoantennas~\cite{KrasnokOE}, photonic topological insulators~\cite{Slobozhanyuk2015}, broadband perfect reflectors~\cite{Krishnamurthy_13, Valentine2014}, waveguides~\cite{Savelev2014_1}, cloacking~\cite{gaillot2008all, cloaking2015all}, harmonics generation~\cite{ShcherbakovNL2014}, wave-front engineering and dispersion control~\cite{Staude_15}.

Such magnetic optical response originates from the circular displacement currents excited inside the nanoparticle by incident light. This opens the possibility of interference between magnetic and electric modes inside the dielectric nanoparticle at some wavelength. One of the most remarkable effects based on this concept is formation of the so-called Huygens source, scattering forward the whole energy~\cite{kerker1983}, while for another wavelength range, the nanoparticle can scatter incident light in almost completely backward direction~\cite{krasnok2011huygens, Lukyanchuk13}. Therefore, manipulation by both electric and magnetic resonances paves the way for effective tuning of the dielectric nanoparticle scattering in the optical range. The spectral positions of the electric and magnetic dipole resonances depend on the dielectric particle geometry and ambient conditions~\cite{evlyukhin2010, Brener_12, Evlyukhin:NL:2012, EvlyukhinSciRep2014, yang2014all, Staude_15}. Another approach for the resonances tuning is to change dielectric permittivity of the particle, which was achieved by means of annealing of amorphous silicon nanoparticles~\cite{chichkov2014NatCom}.

\begin{figure}[!t]
\includegraphics[width=0.45\textwidth]{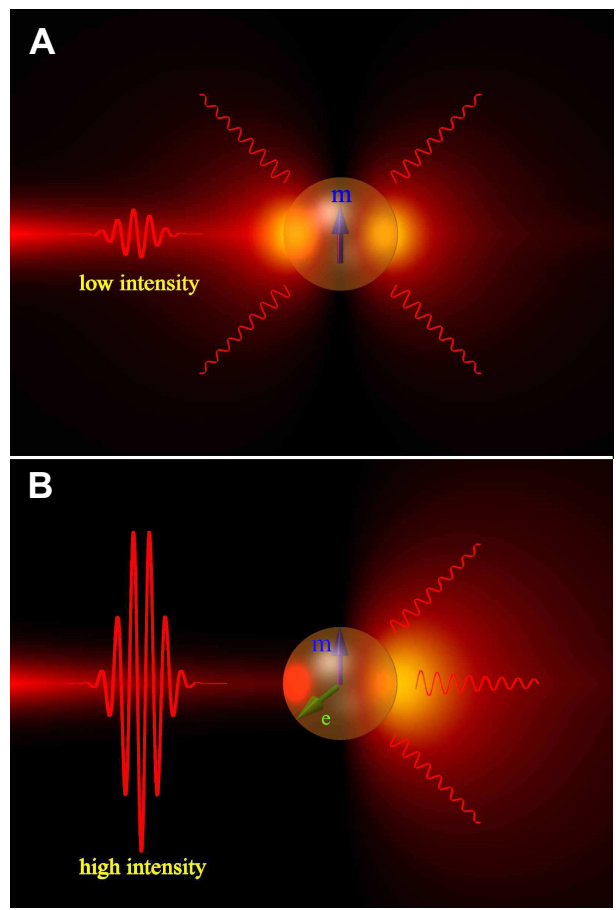}
\caption{A schematic illustration of ultrafast scattering manipulation by a single weak (\textbf{A}) and intense (\textbf{B}) femtosecond laser pulse. Intense femtosecond laser pulse switch-on a Huygens source regime, when incident light is scattered in forward direction.}\label{Fig1}
\end{figure}

However, modern technologies require fast, large, and reversible modulation of optical response of ultracompact functional structures. For this purpose, different types of optical nonlinearities both in metallic ~\cite{zayats2012nonlinear} and dielectric structures~\cite{LPR2015review} have been utilized such as Kerr-type nonlinearities~\cite{zhou2010analytical, Noskov12, lapshina2012nanoradar, abb2014hotspot}, free carriers generation~\cite{nature2004switch, dani2009subpicosecond, large2010} and variation of their temperature~\cite{zayats2011fsnanorods}, as well as relatively slow thermal nonlinearity~\cite{notomi2005optical}. Since plasmonics has high inherent losses and photonic crystals are much larger than the wavelength, it is more desirable to use the "magnetic light" concept, dealing with both low-loss and subwavelength structures. Moreover, nonlinear manipulating by the scattering properties of a nanostructure via its magnetic response gives an additional efficient tool for ultrafast all-optical switching and light modulation.

In this work we propose a novel concept of ultrafast manipulation of the electric and magnetic responses of a high-index dielectric nanoparticle, employing its ultrafast photoexcitation. The existence of the magnetic response makes it possible to switch-on the Huygens source (with cardioid scattering diagram) in the irradiated nanoparticle on femtosecond time scale (Fig.~\ref{Fig1}), enabling to design Huygens metasurface with ultrafast laser-induced transparency. In our experimental study, we prove that the novel concept "ultrafast magnetic light" can be realized in non-destructive regime of light-mater interaction.

\textbf{Ultrafast silicon photoexcitation}--Femtosecond (fs) laser pulses are known to provide strong photo-induced electronic excitation of free carriers in diverse materials (owing to usually low electronic heat capacities), which is accompanied by dramatic variation of their optical characteristics on the timescale of the laser pumping pulse ($<$ 100 fs). Such almost prompt, fs-laser induced optical tunability appears to be much broader for (semi)insulating materials with very minor initial carrier concentrations, extending in a sub-ablative regime in the visible range, e.g., in terms of dielectric permittivity ($\varepsilon$) from large positive (${\rm Re}(\varepsilon)\simeq+10$ for the initial undoped material) to deeply negative (${\rm Re}(\varepsilon)\simeq-10$ for its strongly-excited, metallized surface layer) values (the so-called insulator-metal transition~\cite{Mazur95}). In detail, such ultrafast transient modulation of optical dielectric permittivity in semiconductors and dielectrics is related to transient variation of free-carrier (electron-hole plasma, EHP) density ($\rho_{\rm eh}$) through its basic intraband and interband contributions~\cite{Hirlimann83, Antonetti84, Downer90, Linde2000, Bonse2006, Mazur95, Kudryashov2002, Kudryashov2012, ionin2014electron}. Specifically, an optically driven insulator-conductor transition occurs in such materials, when the fs-laser fluence-dependent EHP frequency passes through the probe frequency, which becomes evident as subsequent reflectivity increase for stronger ionized materials through their intraband electronic transitions~\cite{Hirlimann83, Antonetti84, Downer90}. Simultaneously, ultrafast transient optical modulation for silicon~\cite{Kudryashov2002, ionin2014electron} and other materials~\cite{Mazur95} is additionally enhanced due to a strong prompt EHP-driven isotropic renormalization of their direct bandgap, resulting in drastic enhancement of interband transitions and corresponding red spectral shift of the optical dielectric permittivity~\cite{Mazur95}.

\begin{figure}[!t]
\includegraphics[width=0.5\textwidth]{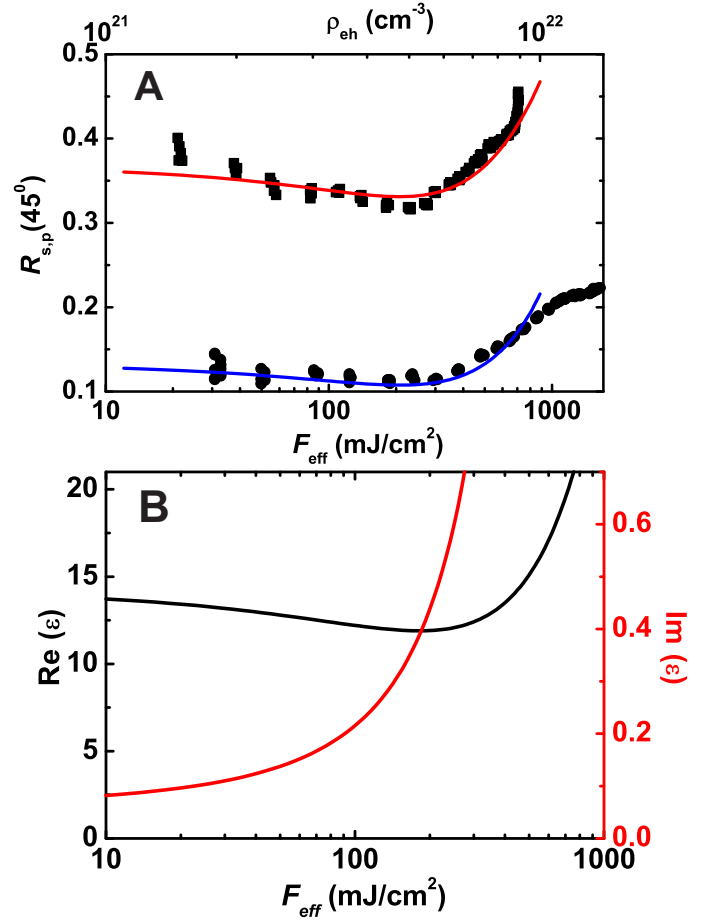}
\caption{(\textbf{A}) Experimental pump self-reflectivity dependences $R_{\rm s,p}(45^\circ,F_{\rm eff})$ (dark squares and circles, respectively) with their corresponding fitting red and blue model curves $R_{\rm s,p}(45^\circ,\rho_{\rm eh})$. (\textbf{B}) Derived dependences of real (black curve) and imaginary (red curve) components of optical dielectric permittivity on $F_{\rm eff}$ at the 800-nm pump wavelength.}\label{Fig2}
\end{figure}

In this work a realistic dependence of the dielectric permittivity for photo-excited silicon versus incident fs-laser fluence at 800-nm laser wavelength was obtained through measurements of single-shot fs-laser pump self-reflectivity (\textit{R}) from smooth Si surface at its \textit{s}- $(R_s(45^\circ))$ and \textit{p}- $(R_p(45^\circ))$ polarizations at the $45^\circ$-incidence angle and variable effective (absorbed) laser fluences $F_{\rm eff}=(1-R_{\rm s,p}(45^\circ,F))$$\cdot$\textit{F}, where \textit{F} is the incident fluence (see \textit{Supplementary materials} for details). Fitting of the experimental dependencies of reflection coefficients $R_{\rm s,p}(45^\circ,F_{\rm eff})$ on fluence using a model transient dielectric permittivity for photo-excited silicon. Commonly, it can be written as a function of $\rho_{\rm eh}$ as a sum of interband- and intraband-transition based terms~\cite{Kudryashov2012, ionin2014electron, Linde2000}:
\begin{eqnarray}
\varepsilon(\omega,\rho_{\rm eh})=\varepsilon_{\rm IB}(\omega^*)\left(1-\frac{\rho_{\rm eh}}{\rho_{\rm bf}}\right)-\nonumber\\
-\frac{\omega^2_{\rm pl}(\rho_{\rm eh})}{\omega^2+1/(\tau^2_e(\rho_{\rm eh}))}\left(1-\frac{i}{\omega\tau_e(\rho_{eh})}\right)
\end{eqnarray}
where the prompt $\rho_{\rm eh}$-dependent bandgap shrinkage effect on interband transitions is accounted by the spectral dielectric permittivity dependence with the effective photon frequency $\omega^{*}= \omega+\Theta\rho_{\rm eh}/\rho_{\rm bgr}$ with the factor $\Theta$, the characteristic renormalization EHP density ($\rho_{\rm bgr}$), the characteristic band capacity ($\rho_{\rm bf}$) of the specific photo-excited regions of the first Brillouine zone in the k-space, the EHP frequency $\omega_{\rm pl}$ and electronic damping time $\tau_e$ at the pump frequency $\omega$ (for details of calculations, see \textit{Supplementary materials}). The resulting model $\rho_{\rm eh}$-dependent pump reflectivities $R_{\rm s,p}(45^\circ,\rho_{\rm eh})$ at the 800-nm pump wavelength were calculated, using a common Fresnel formula and accounting for all basic effects (bandgap renormalization, band filling, EHP screening of the ion core potential~\cite{Mazur95, ionin2014electron}), fitting well the experimental reflectivity dependences $R_{\rm s,p}(45^\circ,F_{\rm eff})$ in Fig.~\ref{Fig2}A with the characteristic initial dip and the following rise. The initial dip implies the predominant intraband (Drude) contribution to the optical dielectric permittivity, while the succeeding reflectivity rise indicates the presumably interband (red spectral shifting) contribution in the normal dispersion region at the "red" shoulder of the $E_1$-band of silicon~\cite{palik}. Such reasonable fitting in Fig.~\ref{Fig2}A provides an important relationship between magnitudes $\rho_{\rm eh}$ and $F_{\rm eff}$ in the region, covering the reflectivity dip and rise, which were used to plot the derived optical dielectric permittivity versus $F_{\rm eff}$ (Fig.~\ref{Fig2}B). Its real and imaginary components at the pump frequency exhibited versus $\rho_{\rm eh}$ the non-monotonously changing -- decreasing and then increasing -- negative real part and monotonously increasing imaginary part  (Fig.~\ref{Fig2}B), respectively. Such dropping ${\rm Re}(\varepsilon)$ and rising ${\rm Im}(\varepsilon)$ magnitudes are consistent with the increasing intraband contribution at the monotonously rising EHP density, while the following rise represents characteristic red-shift changes of these quantities across absorption bands in strongly photo-excited semiconductors due to their prompt $\rho_{\rm eh}$-dependent bandgap renormalization~\cite{Mazur95, Kudryashov2002, ionin2014electron}.

\textbf{Ultrafast photoexcitation of a silicon nanoparticle}--The observed large, ultrafast changes of optical dielectric permittivity properties in silicon at fluences below its ablation threshold (for spallation under these experimental conditions, \textit{F}$_{\rm eff}$~$\approx$~0.3~J/cm$^2$~\cite{ionin2013Si}) can significantly alter optical response of a silicon nanoparticle, supporting a magnetic Mie-type resonance. Basing on the extracted dielectric permittivity values of photoexcited silicon, we study ultrafast dynamics of scattering properties (cross section and diagram) of a silicon nanoparticle near its magnetic resonance by means of full-wave numerical simulations carried out in CST Microwave Studio.

We numerically analyzed optical properties of a spherical (the diameter \textit{D} $\equiv$ 2R = 210~nm) silicon particle with its dilectric permittivity, depending on laser fluence as shown in Fig.~\ref{Fig1}B. The chosen nanoparticle diameter corresponds to excitation of a magnetic dipole Mie-type resonance in the vicinity of the femtosecond laser wavelength $\lambda$ $\approx$ 800~nm. Its extinction cross-section and scattering diagram are well-known to be rather spectrally sensitive in the vicinity of the magnetic resonance~\cite{KrasnokOE, Lukyanchuk13}. In particular, at some wavelengths, where magnetic and electric dipoles induced in the nanoparticle are almost equal and oscillate in phase, the silicon nanoparticle works as a Huygens source with suppressed backward scattering~\cite{KrasnokOE,Lukyanchuk13}. On the other hand, the scattering diagram can be tuned to the regime of suppressed forward scattering ("reverse" Huygens source), when magnetic and electric dipoles oscillate with phase difference of $\pi$/2~\cite{KrasnokOE,Lukyanchuk13}.

\begin{figure}[!t]
\includegraphics[width=0.48\textwidth]{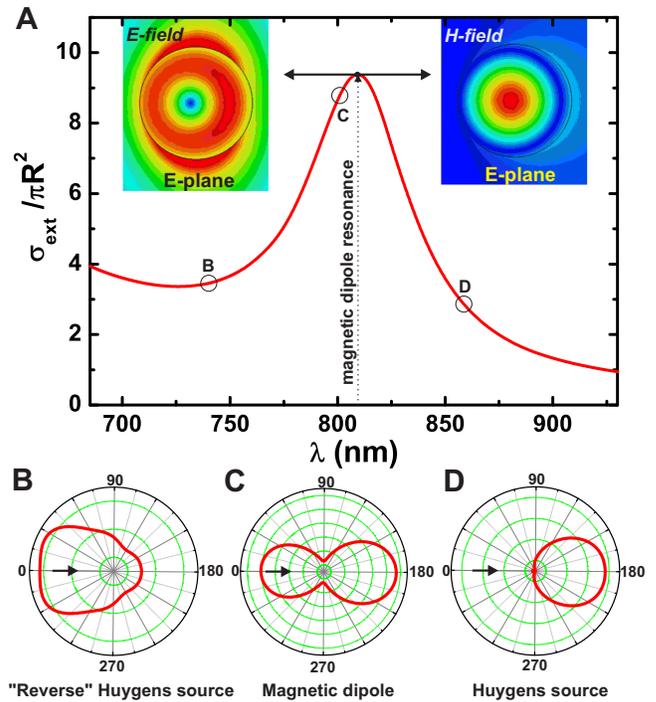}
\caption{Numerically calculated extinction spectrum of a silicon sphere of a radius \textit{D} = 210~nm (\textbf{A}). Insets: numerically calculated electric and magnetic fields distributions near the silicon sphere. Scattering diagrams of the silicon sphere at $\lambda$ = 740~nm (\textbf{B}), 800~nm (\textbf{C}) and 860~nm (\textbf{D}). Black arrows in (\textbf{B}--\textbf{D}) indicate direction of light incidence.}\label{Fig3}
\end{figure}

\begin{figure*}
\includegraphics[width=0.9\textwidth]{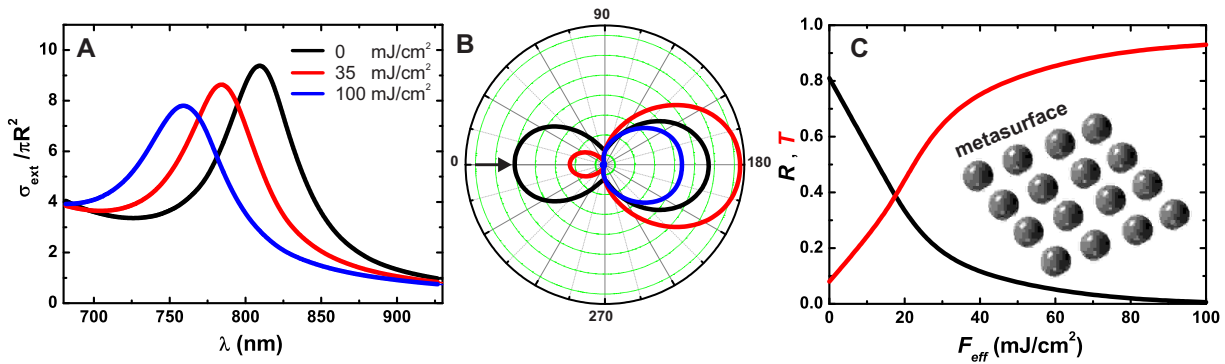}
\caption{Numerically calculated normalized extinction spectra (\textbf{A}) and scattering diagram (\textbf{B}) of the 210-nm large silicon particle irradiated at $\lambda$ = 800~nm and $F_{\rm eff}$ = 0 (black curve), 35 mJ/cm$^2$ (red curve), and 100 mJ/cm$^2$ (blue curve). (\textbf{C}) Numerically calculated dependences of reflectance (black curve) and transmittance (red curve) through array of such 210-nm silicon spheres with the period of 600~nm on effective fluence $F_{\rm eff}$.}\label{Fig4}
\end{figure*}

The results of numerical simulations of extinction and scattering properties of the silicon nanoparticle in the vicinity of the magnetic dipole resonance are shown in Fig.~\ref{Fig3}. The calculated distributions of electric and magnetic fields prove emerging a magnetic dipole moment in this range (Fig.~\ref{Fig3}A). Since both extinction spectra (Fig.~\ref{Fig3}A) and scattering diagram (Fig.~\ref{Fig3}B-D) are strongly wavelength-dependent, it is possible to tune these two parameters by varying either the nanoparticle size or its dielectric permittivity.

In order to considerably tune silicon nanoparticle optical properties near the magnetic dipole resonance, it is necessarily to have time of the mode formation shorter than the laser pulse duration. For \textit{D} = 210~nm silicon nanoparticle, the magnetic resonance mode has approximately 50-nm the full-width at half-maximum, and, therefore, the \textit{Q}-factor is about 16, corresponding to the mode formation time in such open resonator of about 7 fs. Since the typical laser pulse duration of the most commercial femtosecond systems is more than 30 fs and optical changes are stabilized at the middle of the fs-pulse, the character of interaction between the nanoparticle and the most part of the femtosecond pulse is significantly changed in comparison with the case of nanoparticle irradiation by low intensity. Therefore, this concept can be applied for effective ultrafast nanophotonic devices (switchers, modulators, etc.).

To simulate numerically changes in optical properties of photoexcited silicon sphere with the diameter of \textit{D} = 210~nm at the wavelength $\lambda$ = 800~nm, we use the derived above dependencies of ${\rm Re}(\varepsilon)$ and ${\rm Im}(\varepsilon)$ on absorbed laser fluence $F_{\rm eff}$ (Fig.~\ref{Fig2}B). The considered range of absorbed fluences $F_{\rm eff}$ $<$ 100 mJ/cm$^2$ corresponds to non-ablative regime of the laser-particle interaction, but laser fluences are still high enough to generate rather dense EHP ($\rho_{\rm eh}$$\approx$1$\times$10$^{21}$ cm$^{-3}$) for efficient switching of the optical properties of the nanoparticle. As was mentioned before, the scattering diagram of the 210-nm large nanoparticle is almost symmetric at $\lambda=800$~nm and $F_{\rm eff}\approx0$ (Fig.~\ref{Fig3}C), while the extinction cross section ($\sigma_{\rm ext}$) normalized on $\pi$R$^2$ has rather high value of about 9 (see Figs.~\ref{Fig3}A and ~\ref{Fig4}A). The latter parameter is changed almost by three times with fluence increasing up to $F_{\rm eff}$ = 100 mJ/cm$^2$ at the fixed wavelength 800~nm, owing to the strong shift of the peak position of the extinction spectrum (Fig.~\ref{Fig4}A). Its scattering diagram appears to be also very sensitive to the corresponding changes of the dielectric permittivity. For \textit{D} = 210~nm at $\lambda\approx810$~nm, the transition from the typical dipole scattering diagram to the Huygens source forward scattering is observed in the fluence range $F_{\rm eff}=0-100$ mJ/cm$^2$ (Fig.~\ref{Fig4}B).

Since the scattering cross section is reduced and backward scattering is suppressed, almost all energy of the incident light is transmitted through the nanoparticle at negligible reflection. It means that a metasurface based on such elements should demonstrate ultrafast switching from a reflecting to a non-reflecting state. In order to propose the design for the ultrafast switching metasurface, transmission and reflection are numerically calculated for a infinite periodic array of silicon spheres with the diameter \textit{D} = 210~nm and period of 600~nm. The resulting dependences \textit{R}($F_{\rm eff}$) and \textit{T}($F_{\rm eff}$) for the 800-nm wavelength demonstrate significant changes even at $F_{\rm eff}=5$~mJ/cm$^2$ where transmission is increased by more than 100$\%$ (Fig.~\ref{Fig4}C). Taking into account 10-fold average intensity enhancement within each nanoparticle (Fig.~\ref{Fig3}A), the estimated incident fluence, corresponding for 100$\%$ increasing of transmission, is about $F \approx$ 0.5 mJ/cm$^{2}$. For excitation using a diffraction-limited spot of around 1 $\mu$m$^{2}$ area, the above fluence gives a switching energy of 5 pJ, being comparable with microphotonic ring resonators (25 pJ)~\cite{nature2004switch} and different plasmonic nanostructures (7 -- 20 pJ)~\cite{dani2009subpicosecond, large2010, zayats2011fsnanorods}. Moreover, despite the increase of Im($\varepsilon$) with fluence (Fig.~\ref{Fig2}B), the transmission of the silicon nanoparticle-based metasurface is increased by more than 1000$\%$ at $F_{\rm eff}=100$ mJ/cm$^2$ (Fig.~\ref{Fig4}C) due to switching into the Huygens source scattering regime. Therefore, the proposed concept of "ultrafast magnetic light", dealing with both extinction cross section and scattering diagram control, provides low-intensity nonlinear light manipulation by means of ultimately simple nanoobject -- a silicon nanosphere.

\textbf{Conclusion}--In summary, ultrafast generation of dense electron-hole plasma in a dielectric nanoparticle, supporting a magnetic dipole resonance in the optical range, paves the way for the novel concept of "ultrafast magnetic light" based not only on tuning of extinction cross section of the nanoparticle, but also on tuning of its scattering diagram. In the frame of this concept, the possibility of ultrafast switching on a Huygens source in a silicon nanoparticle photoexcited by a femtosecond laser pulse is shown, enabling high-efficient ultrafast light manipulation. Tuning its dielectric permitivity via fs-laser photoexcitation will open a novel class of ultrafast devices, based on the diversity of "magnetic light" effects.

\textbf{Acknowledgments}--This work was financially supported by Russian Science Foundation (Grant 15-19-00172). The authors are thankful to Alexander Poddubny and Mihail Petrov for discussions.

\bibliographystyle{apsrev4-1}
\bibliography{Ultrafast}

\textbf{Supplementary materials for "Ultrafast Magnetic Light"}

\textbf{\textit{Experiment on ultrafast silicon photoexcitation}}

In the experiments we used a commercial 0.45-mm thick, atomically flat undoped silicon Si(100) wafer with a few nanometer-thick native oxide layer, arranged on a three-dimensional motorized translation micro-stage under PC control and raster-moved from laser shot to shot to expose its fresh surface spots. Its single-shot (one laser pulse per surface spot) laser irradiation at a repetition rate of 10 Hz was provided by single IR (800~nm) Ti: sapphire laser pulses in TEM$_{\rm 00}$ mode with a FWHM (full-width at half-maximum) duration $\tau_p\approx100$~fs and pulse energies up to 1.5~mJ, using a triggered electro-mechanical shutter. The pump pulse energy was reduced in these experiments, using a combination of a half-wave plate and a Glan-prism polarizer, to the minimum level slightly above $E_{\rm max}\approx0.2$~mJ to avoid the laser beam degradation due to its self-focusing in air and air plasma scattering/refraction (the critical power at the wavelength $\approx3$~GW~\cite{Mysyrowicz2007}, i.e., $\approx0.3$~mJ for the 100-fs laser pulses).

\begin{figure}[b!]
\centering\includegraphics[width=0.5\textwidth]{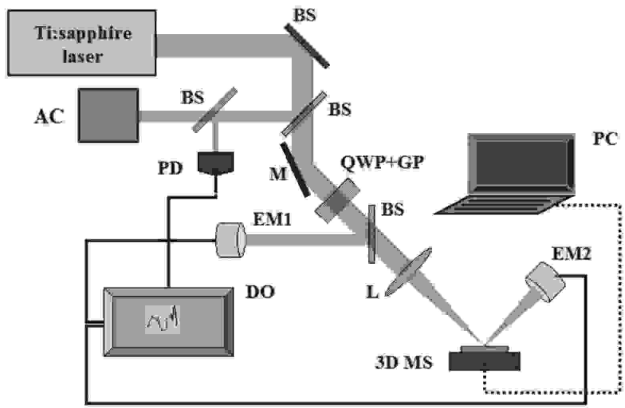}
\caption{Experimental setup: BS--beam splitter, AC--autocorrelator, QPW+GPP--energy attenuating combination of a quarter-wave plate and a Glan-prism polarizer, EM1,2--thermocouple energy meters, M--mirror, L--focusing silica lenses, DO--digital oscilloscope, 3D-MS--three-dimensional motorized micro-stage, PC--laptop for data acquisition and hardware control.}\label{Fig5}
\end{figure}

The fs-laser pump pulses were focused in \textit{s}-- and \textit{p}--polarizations at the angle of $45^\circ$ by a silica lens (focal distance f=500~mm) into a small focal spot (the Gaussian main ellipse 1/e-radii -- $\oslash_{\rm 1/e,x}\approx0.10$~mm and $\oslash_{\rm 1/e,y}\approx0.05$~mm) on the surface of the Si sample. Energies of of incident pump laser pulses $E_{\rm i}\leq E_{\rm max}$ were monitored by means of a pyroelectric energy meter (OPHIR), respectively, while energies $E_{\rm R,i}$ of corresponding reflected pump laser pulses were simultaneously measured 3-5 times for each pulse energy over fresh surface spots by another pyroelectric energy meter (OPHIR), providing the pulse energy dependence of the pump self-reflectance $R_{\rm s,p}(45^\circ,E_{\rm i})$.

In the experimental arrangement of incident and reflected energy measurements ($E_{\rm i}$ and $E_{\rm s,p}$, respectively), the measured self-reflectivity values $R_{\rm s,p}(45^\circ,E_{\rm i})$ represent their true $R_{\rm s,p}(45^\circ,I_{\rm i})$, temporally averaged over the 100-fs wide pump profile \textit{I}(t) and over the laser spot laser spot with the spatial distribution \textit{I}(x,y) characterized by the peak magnitude $I_{\rm i}$
\begin{equation}\label{S1}
R_{\rm s,p}(45^\circ,E_{\rm i})=\frac{E_{\rm s,p}}{E_{\rm i}}=\frac{\iiint R_{\rm s,p}(45^\circ,I_{\rm i})I_i(x,y,t)dxdydt}{\iiint I_{\rm i}(x,y,t)dxdydt},
\end{equation}
where \textit{x,y} are the spatial surface coordinates. The temporal averaging hides the unknown dynamics of EHP in the photo-excited silicon, while the latter spatial averaging can readily be lifted up via a common iterative deconvolution procedure~\cite{Mazur95, Kudryashov2002, Kudryashov2012}, employing $R_{\rm s,p}(45^\circ,E_i)$ values measured at lower energies $E_i$ (peak fluences $F_{\rm i}$) ~\cite{ionin2014electron}.
\begin{eqnarray}\label{S2}
&R_{\rm s,p}(45^\circ,F_i)= \nonumber\\
&\dfrac{R_{\rm s,p}(45^\circ,E_i)-\sum\limits_{k=1}^{i-1}R_{\rm s,p}(45^\circ,F_k)P_k(E_i,F_k,F_{k-1})}{P_i(E_i,F_i,F_{i-1})},
\end{eqnarray}
where local reflectivity values across the laser spot are assumed to be constant in each range $F_{k-1}\div F_k$ and are weighted with their statistic weights $P_k(E_i, F_k, F_{k-1})$, which are equal at each pulse energy $E_i$ to the partial energies in the ring-like squares $\Delta S_k(E_i, F_k, F_{k-1})$ with the boundary local fluences $F_{k-1}$ and $F_k$
{\begin{eqnarray}\label{S3}
&P_k(E_i,F_k,F_{k-1})=\dfrac{\Delta E_k(E_i,F_k,F_{k-1})}{E_i}=\frac{F_k \Delta S_k(E_i,F_k,F_{k-1})}{E_i},\nonumber\\
&\dfrac{\sum\limits_{k=1}^{i}F_k\Delta S_k(E_i,F_k,F_{k-1})}{E_{i}}=1
\end{eqnarray}
}

The spatial deconvolution procedure starts from the lowest experimental $E_1$($F_1$) magnitude, where the dependences $R_{\rm s,p}(45^\circ,E_i)$ exhibit their almost unperturbed reflectivity values~\cite{palik}, and proceeds eventually to higher fluences. In both the cases, the incident fluence calibration, providing the peak surface fluence \textit{F} up to 1.5 J/cm$^2$ as a function of the incident energy and the abovementioned Gaussian beam parameters, was performed by means of an optical microscope to measure main radii of the resulting ablative surface craters.

\textbf{\textit{Details on modeling of ultrafast silicon photoexcitation}}

During modeling of ultrafast transient optics of photoexcited silicon with the help of Eq.~1 in the main text, important instantaneous electronic bandgap renormalization and screening effects were taken into account for the first time. Also, we used the following parameters: the characteristic renormalization EHP density $\rho_{\rm bgr}\approx1\times10^{22}~cm^{-3}$~\cite{ionin2014electron}, being typically about 5$\%$ of the total valence electron density $(\approx2\times10^{23}$~cm$^{-3}$ in Si) to provide the ultimate 50$\%$ electronic direct bandgap renormalization~\cite{Louie2004}, i.e., $\hbar\Theta\approx$ 1.7~eV of the effective minimal gap $\approx3.4$~eV in silicon~\cite{palik, Dargys_book}, while $\rho_{\rm bf}$ is the characteristic band capacity of the specific photo-excited regions of the first Brillouine zone in the k-space (e.g., $\rho_{\rm bf}(L)\approx4\times10^{21}$~cm$^{-3}$ for L-valleys and $\rho_{\rm bf}(X)\approx 4.5\times10^{22}$ cm$^{-3}$ for X-valleys in Si), affecting interband transitions via the band-filling effect~\cite{Mazur95, ionin2014electron, Downer90, Linde2000}. The bulk EHP frequency $\omega_{\rm pl}$ is defined as

\begin{equation}\label{S4}
\omega_{\rm pl}^{2}(\rho_{\rm eh})=\frac{\rho_{\rm eh}e^2}{\varepsilon_0\varepsilon_{\rm hf}(\rho_{\rm eh})m^{*}_{\rm opt}(\rho_{\rm eh})}
\end{equation}
where the effective optical (e-h pair) mass $m^{*}_{\rm opt}\approx0.14$$m_{\rm e}$ in L-valleys or 0.19 in X-valleys~\cite{ionin2014electron, Antonetti84, Linde2000, Grigoropoulos2002, Dargys_book} is a $\rho_{\rm eh}$-dependent quantity, varying versus transient band filling due to the band dispersion and versus bandgap renormalization~\cite{Cardona2005}. The high-frequency electronic dielectric constant $\varepsilon_{\rm hf}$ was modeled in the form $\varepsilon_{\rm hf}(\rho_{\rm eh})=1+\varepsilon_{\rm hf}(0)\times\exp(-\rho_{\rm eh}/\rho_{\rm scr})$, where the screening density $\rho_{\rm scr}\approx1\times10^{21}$~cm$^{-3}$ was chosen to provide $\varepsilon_{\rm hf}\rightarrow1$ in dense EHP. The electronic damping time $\tau_e$ in dense EHP at the probe frequency $\omega_{\rm pr}$ was taken, similarly to metals, in the random phase approximation as proportional to the inverse bulk EHP frequency $\omega_{\rm pl}^{-1}$~\cite{Lagendijk1995}

\begin{equation}\label{S5}
\tau_e=\left(\frac{128E_{F}^2}{\pi^2\sqrt{3}\omega_{\rm pl}}\right)\frac{1+\exp\left[\frac{\hbar\omega}{k_{B}T_{e}}\right]}{(\pi k_{B}T_{e})^2+(\hbar\omega)^2}
\end{equation}

where $E_{\rm F}\approx$ 1 -- 2~eV is the effective Fermi-level quasi-energy for electrons and holes at $\rho_{\rm eh}<1\times10^{22}$~cm$^{-3}$, $\hbar$ and $k_B$ are the reduced Planck and Boltzmann constants, respectively, and $T_{\rm e}$ is the unified EHP temperature, being a weak function of $\rho_{\rm eh}$~\cite{Yoffa1980}. Here, the latter relationship was evaluated for $\hbar\omega>k_{\rm B}T_{\rm e}$ in the form $\tau_{\rm e}(\rho_{\rm eh})\approx3\times10^{2}/(\omega_{\rm pl}(\rho_{\rm eh})\omega^2)$, accounting multiple carrier scattering paths for the three top valence sub-bands, and multiple X-valleys in the lowest conduction band of silicon.

The resulting $\rho_{\rm eh}$-dependent model oblique-incidence pump reflectivities $R_{\rm s,p}(45^\circ,\rho_{\rm eh})$ calculated, using common Fresnel formulae, fit well the extracted experimental reflectivity dependences $R_{\rm s,p}(45^\circ,F_{\rm eff})$ in Fig.~2A with the characteristic initial dip and the following rise.

\end{document}